\documentclass[aps,prl,reprint]{revtex4-1}

\usepackage{amsmath}
\usepackage{amssymb}
\usepackage{graphicx}
\usepackage{hyperref}
\usepackage{blindtext}

\begin{abstract}
We report with single particle simulations that longitudinal chromatism, a commonly occuring spatio-temporal coupling in ultrashort laser pulses, can have a significant influence in the longitudinal acceleration of electrons via high-power, tightly-focused radially polarized laser beams. This effect can be advantageous, and even more-so when combined with small values of temporal chirp. However, the effect can also be highly destructive when the magnitude and sign of the longitudinal chromatism is not ideal, even at very small magnitudes. This motivates the characterization and understanding of the driving laser pulses, and further study of the influence of similar low-order spatial-temporal couplings on such acceleration.
\end{abstract}
\begin{document}

\title{Influence of longitudinal chromatism on vacuum acceleration by intense radially polarized laser beams}
\author{Spencer W. Jolly}
\email{spencer.jolly@cea.fr}
\affiliation{LIDYL, CEA, CNRS, Universit{\'e} Paris-Saclay, CEA Saclay, 91 191 Gif-sur-Yvette, France}
\date{March 11, 2019}
\maketitle

The acceleration of electrons to relativistic energies in vacuum is possible with a tightly focused, high power radially polarized laser beam (RPLB). The solution of the wave equation in focus produces a purely longitudinal field on axis, with a radial field towards the axis at points slightly off-axis~\cite{salamin06}. This provides the motivation for using these beams to accelerate particles, either from rest or from modest energies, to relativistic energies~\cite{varin05,salamin07}. We simulate the effects of the spatio-temporal coupling longitudinal chromatism (or chromatic focusing) on such acceleration, limited initially to particles on-axis, but also expanded to one specific case including off-axis trajectories. Since longitudinal chromatism affects the field mainly longitudinally, the effects on the acceleration enabled by the strong longitudinal fields of the focused RPLB could be severe.

The most general case of vacuum laser acceleration by an RPLB is that where a test particle begins from rest (or with a moderate inital energy) and an RPLB focuses and overtakes the particle. The longitudinal field in the focal region, when the laser power is large enough, imparts a net kinetic energy on the particle after overtaking. Detailed studies have been done showing especially that with optimization of both the initial position of the test particle and the carrier-offset phase (CEP) of the driving laser, the kinetic energy of the particle is always higher with decreasing duration and/or decreasing focused spot size~\cite{wong10}. These studies included particles beginning at rest and a large range of laser powers showing the suitability of the mechanism. Studies were also done including nonparaxial terms~\cite{marceau12}, off-axis fields~\cite{marceau13-2} and even more complex interactions~\cite{sell14,varin16,wong17} showing that low energy-spread and collimation are indeed possible in a bunch of electrons with finite charge and size. Experimental results of this nature have been achieved in a low density neutral gas achieving 23\,keV energies~\cite{payeur12}, and in a true vacuum, accelerating some electrons in a bunch from 40\,keV up to a maximum of 52\,keV~\cite{carbajo16}.

Because this mechanism requires essentially for the particle to gain enough energy in a single half-cycle so as to not be fully decelerated in a subsequent half-cycle, there is a threshold power below which the particle will gain negligible energy. This has been shown to be related to the normalized beam intensity, and scales with the beam waist ($w_0$) as ${w_0}^4$~\cite{fortin10}. Above this threshold the particle gains significant energy, which is increasing with laser power. Well above the threshold the main limitation is that the accelerated particle inevitably slips out of the position of highest accelerating field since it is always travelling with a subluminal velocity. The maximum electron energy from a transform limited pulse has been shown to scale with the beam power ($P$) as $\sqrt{P}$, with higher intensity pulses approaching this limit~\cite{wong10}. There have been numerical studies showing that it is possible to have on-axis acceleration to high energies with the interaction of a focused radially polarized beam with a plasma mirror~\cite{zaim17}, but this work will focus on the most general case.

Beyond the use of a single Fourier-limited beam producing net acceleration, it may be possible to optimize the interaction with more complex or structured pulses, which we demonstrate in this work. A study was done in the general case using a beam composed of components of two colors with independent CEP~\cite{wong11}. The study showed a drastic increase in single particle accelerated energy with the same total laser power. This was due to the pulse beating within the Rayleigh range and the Gouy phase providing an advantageous situation for the particle to slip less out of the region of high accelerating field. Only certain differences in CEP between the different color pulses produced improved kinetic energies, corresponding to the case where the advantageous phase situation is created after the pulses focus, where the acceleration mainly occurs.

This concept of discrete multi-color fields in focus can be generalized to a specific continuous case, where the different colors of a single broadband driving laser focus to different longitudinal positions. This is known as longitudinal chromatism (LC), which on the collimated beam takes the form of pulse-front curvature (PFC) or a radially varying group delay (for a transform limited pulse). In focus the pulse duration is increased according to the longitudinal separation of the spectral content, resulting in a decreased intensity. This is one of the well known low-order spatio-temporal couplings~\cite{akturk10}. LC can be induced simply by focusing via a chromatic singlet lens~\cite{bor89-1}, by using an afocal doublet made of special glasses to apply PFC on the collimated beam to be focused~\cite{sainte-marie17}, or via a diffractive lens~\cite{froula18} among other methods. We simulate the effect of LC/PFC and chirp on vacuum laser acceleration in the aforementioned general case (related to interaction with a low-density ambient gas). As discussed, PFC and LC are equivalent, but to decouple the focusing geometry and the longitudinal chromatism we use the PFC on the collimated beam as the unique control parameter.

In all of the following simulations we use pulses that have Gaussian spatial and temporal profiles, with $w_0$=4\,$\mu$m and $\tau_0$=10\,fs characteristic widths respectively, at a central wavelength of 800\,nm ($\omega_0$=2.35\,$\times10^{15}$\,rad/s). These parameters both fit within the paraxial approximation. Although it has been shown that even in the case of $w_0$=4\,$\mu$m (5$\lambda_0$) the results of acceleration of off-axis particles including non-paraxial terms can be significantly different than the case without, the on-axis case is still valid~\cite{marceau13-1}. Therefore the simulations presented first should be taken as valid only in the on-axis case, and do not yet provide insight in to acceleration of a beam of electrons with finite extent. Lastly, since the manifestation of the nearfield PFC in focus depends on the focusing geometry, the corresponding near-field width is chosen to be $w_i$=4\,cm with a focal length of f=63\,cm.

The longitudinal field of the focused radial polarized field $E_z$ is modeled with near-field PFC $\alpha$ and group delay dispersion (GDD) $\phi_2$ in the frequency domain as in ~\cite{sainte-marie17} using the proper form for the longitudinal field as in~\cite{wong10}. With $A(\omega)=\exp(-\delta\omega^2/\Delta\omega^2)$, $\Delta\omega=2/\tau_0$, and $\delta\omega=(\omega-\omega_0)$ we have the field around the focus $z=0$ as

\begin{align}
	\hat{E}_z(z,\omega)&=\frac{1}{\Delta\omega}\sqrt{\frac{16P}{\pi\epsilon_0{c}}}\frac{A(\omega)}{z_R\left(1+\left(\frac{z-z_0(\omega)}{z_R}\right)^2\right)}e^{i\phi(z,\omega)} \\
	\phi(z,\omega)&=\Psi_0+2\tan^{-1}\left(\frac{z-z_0(\omega)}{z_R}\right)-\frac{\phi_2\delta\omega^2}{2}-\frac{\omega{z}}{c} ,
\end{align}

\noindent with the Rayleigh length $z_R=2cf^2/\omega_0{w_i}^2$, the frequency dependent focus position due to the LC/PFC $z_0(\omega)=2cf^2\alpha\delta\omega/\omega_0$, the CEP $\Psi_0$, and the Fourier-limited pulse power $P$. In the time domain a positive $\alpha$ corresponds to a positive radial group delay, or the higher frequencies being focused at a higher $z$. The acceleration of the electron is modeled by the relativistic Lorentz force equations

\begin{align}
	\frac{\partial{\beta}}{\partial{t}}&=\frac{-q_e E_z(z,t)(1-\beta^2)^{3/2}}{m_e c} \\
	\frac{\partial{z}}{\partial{t}}&=\beta{c} ,
\end{align}

\noindent with $q_e$ and $m_e$ the elementary charge and electron mass respectively, and $\beta=v/c$. We use a 5th-order Adams-Bashforth finite difference method to have improved accuracy with a fixed step size. Since we only know the field in frequency space, the field $\hat{E}_z(z,\omega)$ must be inverse Fourier transformed to $t$ at each iteration to find the field at the new position $z(t_i)$ and to calculate the next velocity and position. This step significantly increases the computation time compared to the calculations done directly in the time domain (possible only if $\alpha=0$). The simulations are run until the particle energy is no longer changing significantly, generally in the range of 10\,ps for the power levels studied here. Results without any LC/PFC or chirp were compared to those in the literature~\cite{wong10,zaim17} and agreed very well.

\begin{figure}[tb]
	\centering
	\includegraphics[width=84mm]{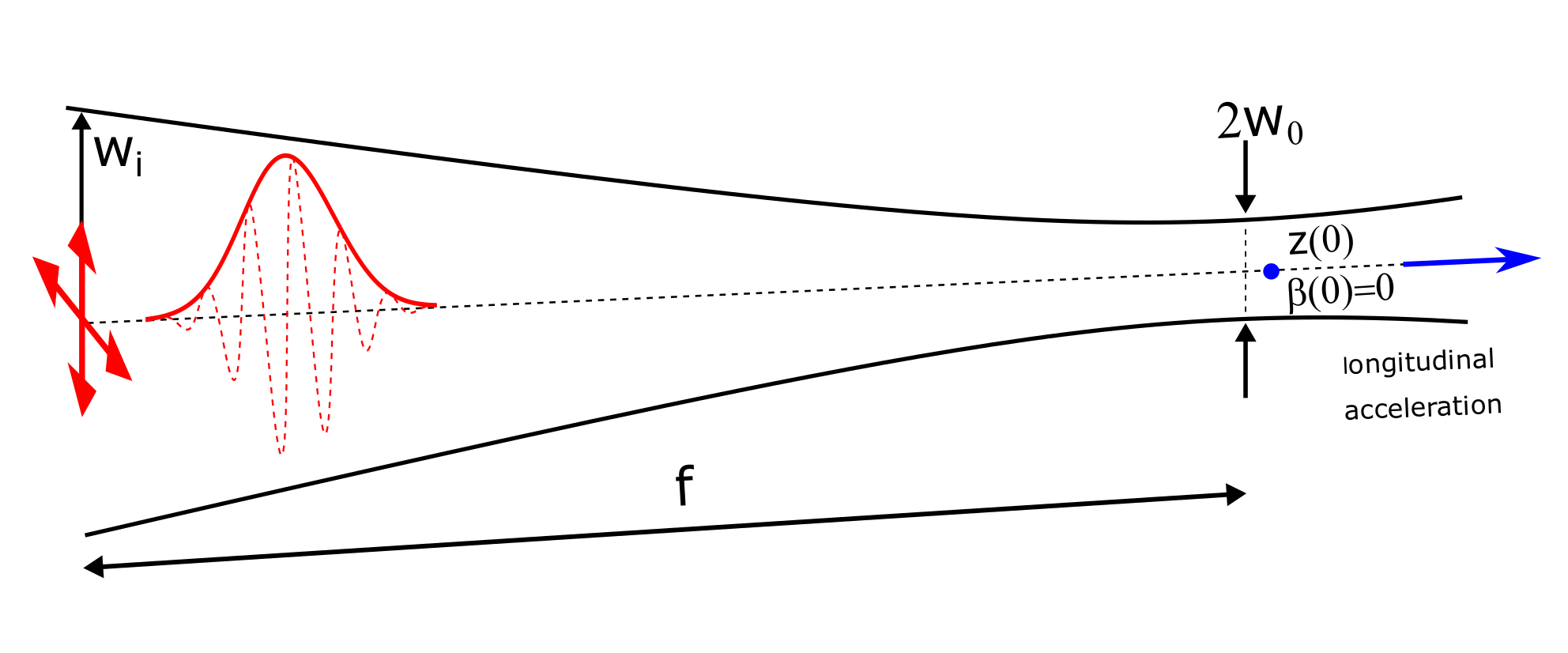}
	\caption{Sketch of the scenario simulated here, with the driving laser pulse having longitudinal chromatism. See the text for more details.}
	\label{fig:ScenariosTech}
\end{figure}

The specific physical scenario is detailed again technically in Fig.~\ref{fig:ScenariosTech}. The simulations start such that the laser pulse is far away from the test particle so as to have no influence, and it propagates across the particle to impart a net kinetic energy. In every simulation --- varying laser power, GDD, and LC/PFC --- we optimize the initial electron position $z(0)$ and laser CEP $\Psi_0$ for the maximum final kinetic energy, with the particles always beginning at rest. This emulates an experimental situation where the particles in a low-density gas would be distributed in space and the driving laser would have tunable CEP.

\begin{figure}[tb]
	\centering
	\includegraphics[width=84mm]{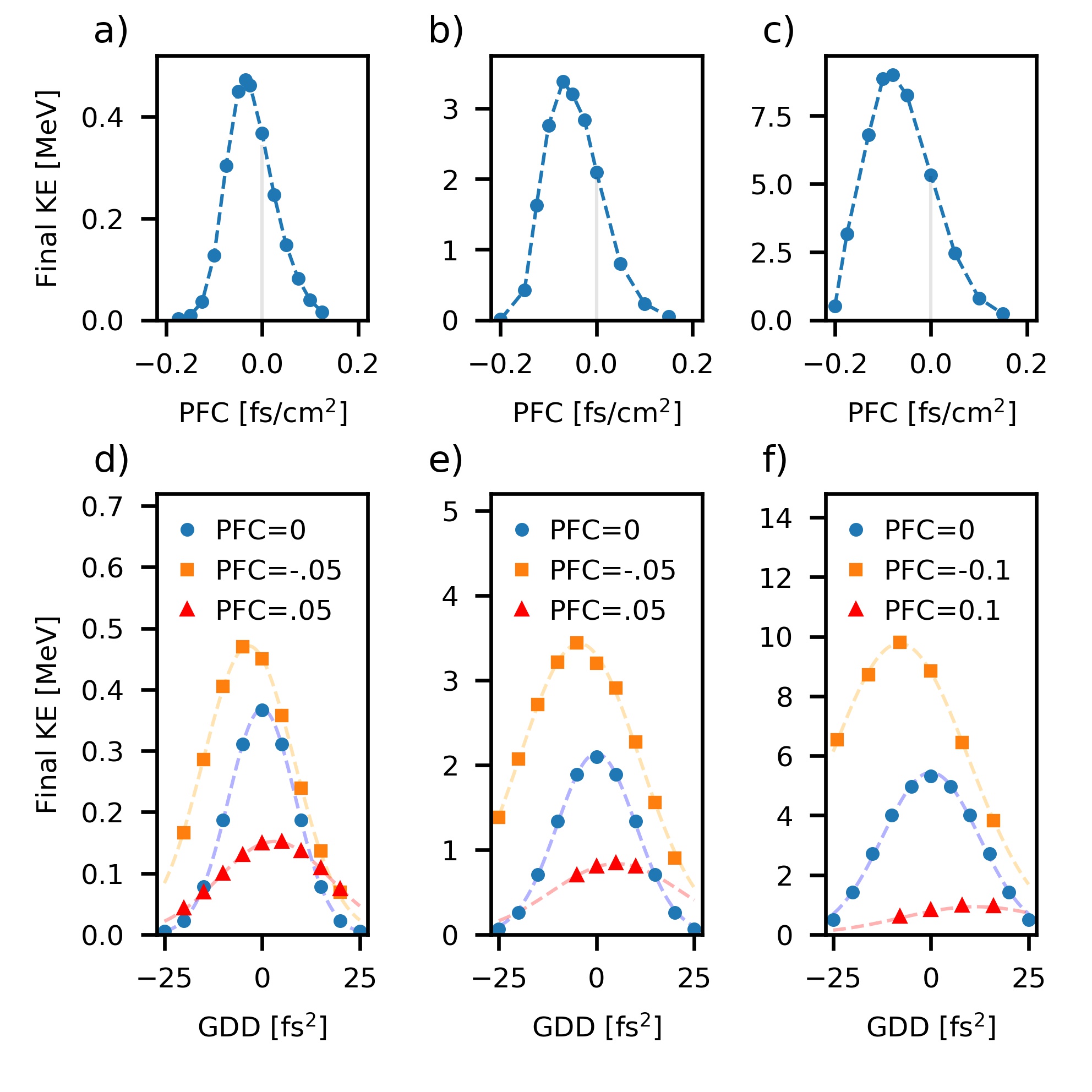}
	\caption{Results of on-axis acceleration with Fourier-limited powers of \textbf{(a,d)} 80, \textbf{(b,e)} 90, and \textbf{(c,f)} 100\,TW. The top row \textbf{(a--c)} is for zero GDD and various amounts of LC/PFC with the dashed lines as guides for the eyes, and the bottom row \textbf{(d--f)} is final kinetic energies for many combinations of LC/PFC and GDD with the dashed lines Gaussian fits. LC/PFC is in units of fs/cm$^2$ always.}
	\label{fig:Scenario1Results1}
\end{figure}

We find that at pulse energies so as to create few-MeV level electrons (Fourier-limited powers of 80, 90, 100\,TW), the addition of LC/PFC has a significant effect on the maximum electron energy after interaction, and the addition of small amounts of chirp can optimize the energy further. This is summarized in Fig.~\ref{fig:Scenario1Results1}. The acceleration decreases for all GDD values when the LC/PFC is zero, matching previous results~\cite{hogan-lamarre15}, showing that the effect can only be optimized with non-zero LC/PFC. Note that for the three power levels studied the magnitude of optimum LC/PFC increases, along with the optimum GDD, and also the relative improvement when adding LC/PFC. The optimum LC/PFC is always negative, but the optimum GDD at any amount of LC/PFC always has the same sign as that LC/PFC.

The striking result of an increased final kinetic energy with this imparted spatio-temporal coupling (and therfore a slightly lower peak intensity) provides direct motivation for the opportunity to optimize the acceleration process via specific fine-tuning of the LC/PFC. However, this decrease at positive values of LC/PFC provides a useful insight into experiments as well. If there is an undiagnosed level of LC/PFC, then the acceleration process could be either enhanced or damped. For example, in the case of acceleration with a Fourier-limited 90\,TW beam a LC/PFC of +0.05\,fs/cm$^2$ results in a decrease in final kinetic energy by a factor of 2.6 (without any chirp). However, on a beam of $w_i$=4\,cm with f=0.63\,m this level of LC/PFC causes a decrease in peak intensity of only 2.5\,\% and is equivalent to a delay of only 0.8\,fs at the edge of the collimated beam. This highlights the sensitivity of the mechanism to LC/PFC. Detailed knowledge of spatio-temporal couplings --- possible with developing full spatio-temporal characterization devices~\cite{pariente16,borot18}, or with diagnostics made specifically for measuring PFC~\cite{bor89-2,li18-2} --- is crucial to not only optimizing the mechanism, but having it perform at a nominal level.

The optimum LC/PFC value is always negative, with all positive values of LC/PFC producing lower net acceleration. This corresponds to longer wavelengths being focused to greater $z$, according to $z_0(\omega)$, meaning that along the acceleration direction the central wavelength is locally increasing within the Rayleigh range. This is a useful way to visualize the effect, analogous to some RF accelerators having increased cavity lengths along the acceleration direction.

In combination with temporal chirp this form of spatio-temporal coupling has been explored to control the velocity of the intensity peak of focused laser pulses, a 'Flying Focus'~\cite{sainte-marie17,froula18}, which was the original motivation for this study. However, the magnitudes of LC/PFC and chirp presented are so small such that the velocity of the intensity peak is not significantly affected, rather the effect on the phase within the Rayleigh range is also important. This is required for any effect on vacuum laser acceleration since both LC/PFC and chirp independently reduce the intensity in focus, and for such acceleration the electric field strength and thus intensity are key parameters.

\begin{figure}[tb]
	\centering
	\includegraphics[width=84mm]{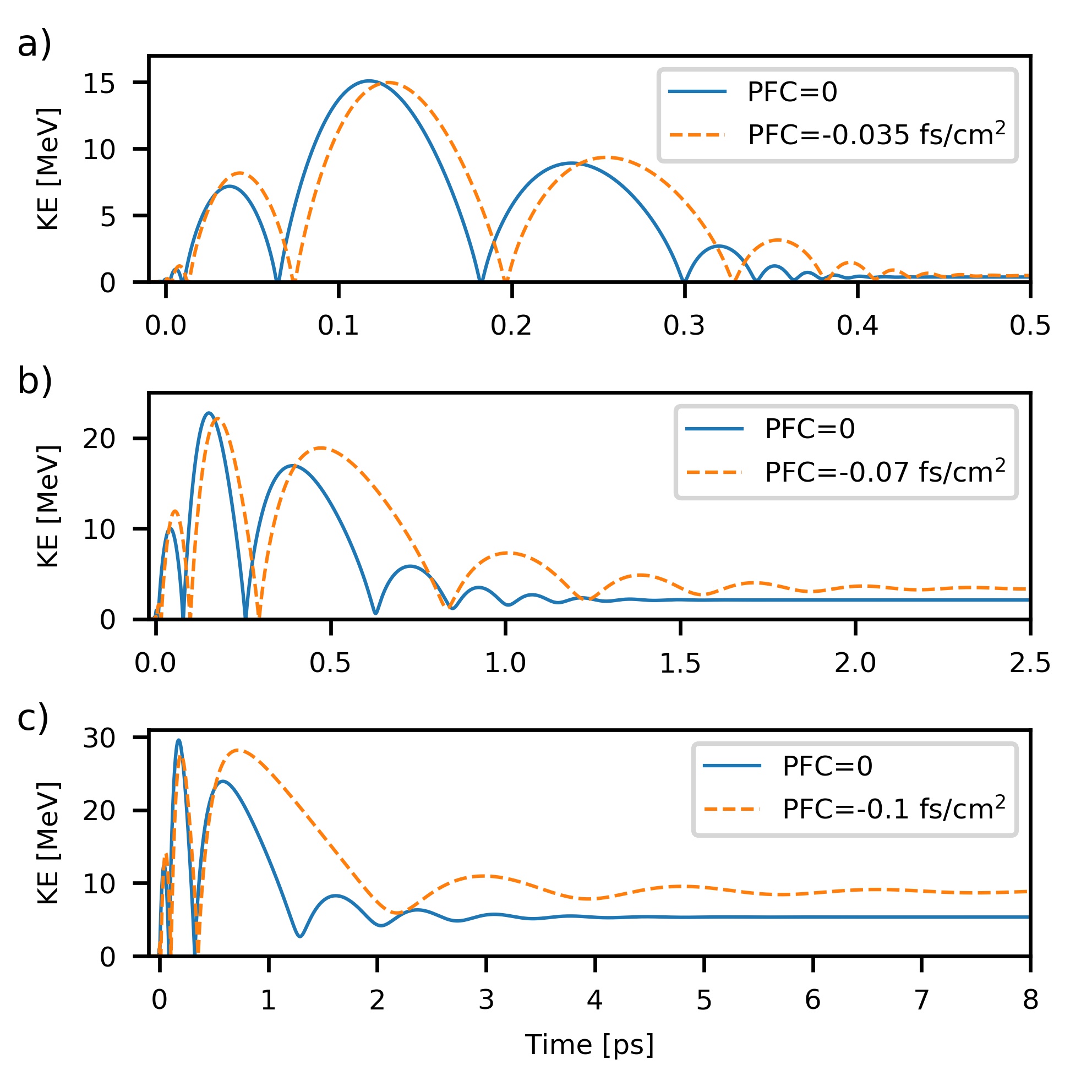}
	\caption{Comparison of trajectories with Fourier-limited powers of \textbf{(a)} 80, \textbf{(b)} 90\,TW, and \textbf{(c)} 100\,TW. Shown are the default cases (PFC=GDD=0) (solid) and the improved cases (dashed).}
	\label{fig:Trajectories}
\end{figure}

To get a better understanding we can also compare the trajectories of representative simulations, seen in Fig.~\ref{fig:Trajectories}. With negative LC/PFC the peak energy reached during the peak accelerating half-cycle is lower, but in subsequent laser cycles the electron is deccelerated less, and accelerated more, leading to a higher final energy. This is sensible, since the peak field is lower in this main accelerating half-cycle, but the LC/PFC has provided an advantageous situation for the region outside of the intensity peak. This is seen most clearly in comparing the dashed trajectory to the solid trajectory in Fig.~\ref{fig:Trajectories}(c) in the time window around 1\,ps.

In order to expand these simulations to the more experimentally relevant case considering electrons beginning off-axis, and therefore not only the $E_z$ component of the fields, we must expand the model to include non-paraxial corrections~\cite{marceau13-1,martens14,favier17}. Using the fields expanded in the small parameter $\epsilon=w_0/z_R=w_i/f=0.0635$ we can calculate very accurately the full fields, now including $E_z$, $E_r$ and $B_{\theta}$ up to terms of order $\epsilon^5$~\cite{salamin06}. To calculate the effect of the LC/PFC on these full fields, we employ the same construction as in Eq.~1 in the frequency domain now with the non-paraxial fields, and inverse Fourier-transform to find the fields in time. Finally, the forces are calculated on the particle, now both in the $z$ and $r$ direction and including the full vector form of the Lorentz force. This formalism was also compared to past work without any LC/PFC~\cite{martens14} and the results agreed well.

\begin{figure}[tb]
	\centering
	\includegraphics[width=84mm]{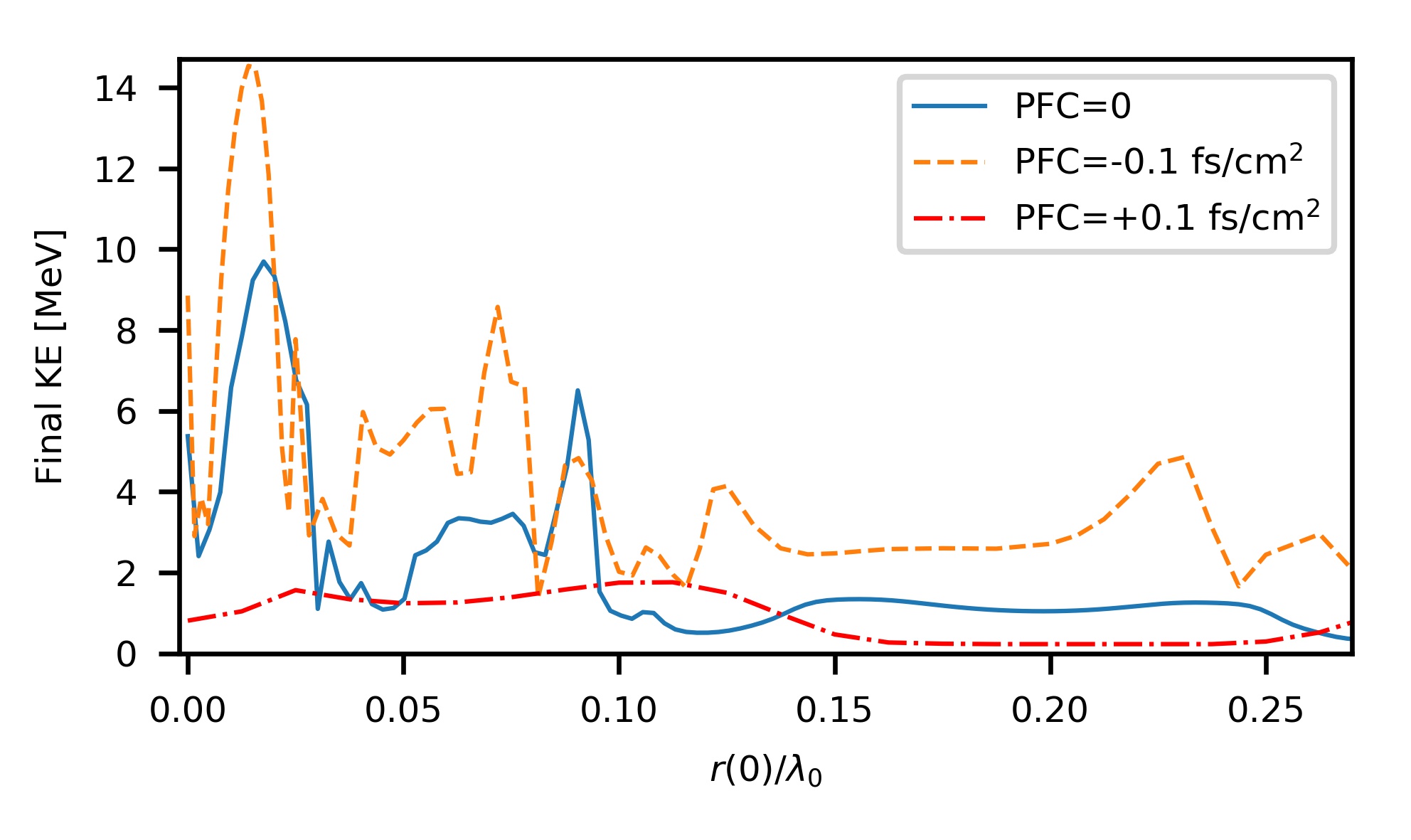}
	\caption{Off-axis acceleration with a Fourier-limited power of 100\,TW. Each case is simulated for the initial electron position $z(0)$ and laser CEP $\Psi_0$ that produces the maximum on-axis acceleration.}
	\label{fig:off-axis}
\end{figure}

The results of off-axis acceleration with non-zero LC/PFC (zero chirp) are shown in Fig.~\ref{fig:off-axis} with a Fourier-limited power of 100\,TW, varying the initial radial position $r(0)$. For each value of LC/PFC the initial electron position $z(0)$ and laser CEP $\Psi_0$ are kept constant, and are set to the values that obtain the maximum on-axis kinetic energy (that is, the same values as for Fig.~\ref{fig:Scenario1Results1}(f)). The main result is that, although the situation becomes significantly more complex, the increase or decrease in final kinetic energy afforded by the PFC is also mostly present in the off-axis acceleration. This can be quantified by the average kinetic energy up to $r(0)/\lambda_0=0.25$, which is larger in the case of $\alpha=-0.1$\,fs/cm$^2$ by a factor of 2.3 compared to $\alpha=0$, and the case of $\alpha=+0.1$\,fs/cm$^2$ lower by a factor of 2.5. Indeed, these off-axis results do not provide insight on electron beam properties such as energy spread and emmitance, with more complete simulations necessary. Nevertheless it is clear that LC/PFC has a strong effect on off-axis acceleration, which in turn strongly influences final parameters of any electron beam with finite duration and transverse extent.

The results in other focusing conditions, with shorter pulse durations, larger driving laser power, or different central frequency are beyond the scope of this work. However, preliminary simulations show that the effect is different at lower pulse durations. Because the effect of LC/PFC on focused intensity is larger as the pulse duration is decreased, the trade-off between improved accelerating phase and decreased intensity is less able to result in an increased final kinetic energy. So it may be that the LC/PFC is purely a detrimental property for other driving laser parameters. Scaling with wavelength is discussed in the literature~\cite{wong10}, and the effect of spatio-temporal distortions on acceleration with developing THz sources is also of potential interest. At higher laser powers, shorter durations, or tighter focusing, a different temporal or spectral profile would need to be used for satisfactory accuracy, along with potentially higher orders of non-paraxial terms.

Experimentally it is possible to compensate for an existing level of LC/PFC~\cite{bahk14}, but such compensation or tuning mechanisms are not commonplace or simple. Additionally, LC/PFC can come from many sources in ultrafast laser systems, so it is difficult to remove completely from the final high-power beam without taking great care. Therefore, practically, the results of these studies do provide an avenue for optimization, but the more relevant result may be that the amount of LC/PFC necessary to spoil the mechanism is very small.

The results in summary mean that the specifically applied spatio-temporal coupling of longitudinal chromatism (or pulse-front curvature in the near-field) can increase the final kinetic energy of a single electron accelerated by a focused ultrashort radially-polarized laser beam. This can be increased further with the addition of small amounts of linear chirp. In the general case of where the initial particle position and the laser CEP are freely varied, this results in almost doubling the energy when accelerated with a 100\,TW beam, confirmed as well in limited off-axis simulations. In every case there is a drastic decrease of the final kinetic energy at values of positive sign, opposite that of the optimum, motivating the characterization of spatio-temporal couplings in any experiment of this kind. Beyond the impact on vacuum laser acceleration presented here, specific combinations of other low- or high-order spatial-temporal couplings may prove useful in particle manipulation or engineering specific laser-material interaction.

\bigskip
\paragraph{Funding.} The work was supported by the fellowship CEA-Eurotalents.2014--2018 (n' PCOFUND-GA-2013-600382) under the Seventh Framework Program (FP7).

\bigskip
\paragraph{Acknowledgement.} The author would like to thank Rapha{\"e}l Lebrun, Antoine Jeandet, and Fabien Qu{\'e}r{\'e} for helpful discussions and comments.

\end{document}